\documentclass{article}
\usepackage{fleqn}
\usepackage{amsmath,amssymb}
\usepackage{verbatim}

\def\stackunder#1#2{\mathrel{\mathop{#2}\limits_{#1}}}

\newcommand{\Je}{J(\varepsilon;\varepsilon_{0})}
\newcommand{\drob}{\frac{1 - p'_{\varepsilon}}
{\varepsilon + p(\varepsilon)}}
\newcommand{\Jeoe}{\int \limits_{\varepsilon_{0}}^{\varepsilon}}

\newcommand{\eopo}{\varepsilon_{0} + p_{0}}
\begin{document}

\begin{center}
{\bf\Large  Magnetohydrodynamic Equations in a Gravitational Field
and Excitation of Magnetohydrodynamic Shock Waves by a
Gravitational Wave}\\
Yu.G.Ignatyev\\
Kazan State Pedagogical University,\\ Mezhlauk str., 1, Kazan
420021, Russia
\end{center}

\begin{abstract}
On the basis of simple principles we derive and investigate the
equations of relativistic plasma magnetohydrodynamics (MHD) in an
arbitrary gra\-vi\-ta\-ti\-onal field. An exact solution
describing the motion of magnetoactive plasma against the
background of the metric of a plane gravitational wave (PGW) with
an arbitrary amplitude is obtained. It is shown that in strong
magnetic fields even a sufficiently small amplitude PGW can create
a shock MHD wave, propagating at a subluminal velocity.
Astrophysical consequences of the anomalous plasma acceleration
are considered.
\end{abstract}

\section{Introduction}
In [1–4] the effect of PGW on plasmalike media was investigated by
the methods of relativistic kinetic theory in the approximation
when the back reaction of matter on the PGW is negligible:
\begin{equation}
\label{1} \varepsilon \ll \omega^{2},
\end{equation}
where $\omega$ is the characteristic frequency of a PGW,
$\varepsilon$ is the matter energy density ($G = \hbar = c = 1$).
These papers have revealed a number of phenomena of interest,
consisting in the induction of longitudinal electric oscillations
in the plasma by PGW. In spite of the strictness of the results
obtained in [1.4], the effects discovered in these papers have
very little to do with the real problem of GW detection. Moreover,
the above calculations show lack of any prospect for GW detectors
based on dynamic excitation of electric oscillations by
gravitational radiation. There are two reasons for that: the
smallness of the ratio $( m^{2} G/e^{2} ) = 10^{-43}$  and the
small relativistic factor $\langle v^2\rangle/c^2$ of standard
plasmalike systems. The GW energy transformation coefficient to
plasmatic oscillations is directly proportional to a product of
these factors.

However, the situation may change radically if strong electric or
magnetic fields are present in the plasma. In Ref. [5], where the
induction of surface currents at a metal-vacuum interface by a PGW
was studied, it was shown that the values of currents thus induced
can be of experimental interest. In [6], on the basis of
relativistic kinetic equations, a set of MHD equations was
obtained, which described the motion of collisionless
magnetoactive plasma in the field of a PGW of an arbitrary
magnitude in a drift approximation and it was shown that, provided
the propagation of the PGW is transversal, there arises a plasma
drift in the PGW propagation direction. The set of equations
obtained in [6] is rather complex and unwieldy: it is a set of
nonlinear partial differential equations. In [7], however, it was
shown that, provided the plasma is originally electroneutral and
uniform, the solution of the above set of equations is strictly
stationary, i.e., it depends only on retarded time. This fact
permits us to substantially simplify the problem and to find its
exact solution, possessing a number of remarkable peculiarities.

\section{The conditions of magnetic field embedding in the plasma}
As pointed out above, in [6], on the basis of a selfconsistent set
of collisionless kinetic equations and the Maxwell equations
(i.e., on the basis of general relati\-vis\-tic Vlasov equations
[8]), a set of MHD equations describing the motion of
magneto\-active plasma in the field of a PGW, was obtained. This
set of equations is obtained in the so-called drift approximation,
i.e., in the first approximation in the small parameter  $\xi$ :
\begin{equation}
\label{2.1} \xi = \frac{\omega}{\omega_{B}} \ll 1 ,
\end{equation}
where $ \omega_{B} = e H/m_{e}c $ is the Larmor frequency.
However, the equations obtained in [6] are applicable only in the
case of a strictly transverse PGW propagation, where the original
magnetic field is perpendicular to the GW propagation
di\-rec\-tion. It is not difficult to verify that if the
conditions of the strict transvesity of PGW propagation are not
met, the equations of [6] violate the energy and momentum
conservation laws. For our purposes it is necessary to consider a
more general case, so in what follows we shall obtain the MHD
equations on the basis of other principles.

It is not difficult to see that a consequence of the MHD equations
from [6] is the magnetic field embedding in the plasma (MFEP).
This reflect the general nature of magnetoactive plasma provided
that the condition (2) is met. Therefore, in order to describe the
motion of the plasma in a drift approximation, it is simpler to
demand at once that the MFEP condition be met. Mathematically this
requirement means a coincidence between the timelike eigenvectors
of the plasma energy-momentum tensor (EMT), $\stackrel {p}{T_{ik}}
$, and that of the electromagnetic field, $\stackrel {f}{T_{ik}}
$, i.e., according to Synge [9] , a coincidence of the dynamic
velocities of the plasma and the electromagnetic field:
\begin{equation}
\label{3}\stackrel{p}{T^{i}_{k}} u^{k} = \varepsilon_{p} u^{i} ,
\end{equation}
\begin{equation}
\label{4}\stackrel{f}{T^{i}_{k}} u^{k} = \varepsilon_{H} u^{i},
\end{equation}
where
\begin{equation}
\label{2.4}(u,u) \stackrel{Df}{=} g_{ik} u^{i} u^{k} = 1.
\end {equation}

We shall consider in this paper the EMT of the plasma as that of a
{\it perfect isotropic fluid}
\begin{equation}
\label{2.5}\stackrel{p}{T^{ik}} = (\varepsilon + p) v^{i} v^{k} -
p g^{ik},
\end{equation}
where
\begin{equation}
\label{2.6}(v,v) = 1 ,
\end{equation}
and $ \varepsilon$ and $p$ are the fluid energy density and
pressure connected by a certain equation of state:
\begin{equation}
\label{2.7}p = p(\varepsilon) .
\end{equation}

Thus $ v^{i}$ is the timelike eigenvector of the plasma EMT
$(v^{i}=u^{i})$, while å is the eigenvalue $\stackrel{p}{T^{ik}}
(\varepsilon_{p}=\varepsilon)$, and there remain the conditions
(\ref{4}):
\begin {equation}
\label{2.8}\stackrel{f}{T^{i}_{k}} v^{k} = \varepsilon_{H} v^{i} .
\end{equation}

Thus (\ref{2.8}) are precisely the conditions of magnetic field
embedding in the plasma (MFEP). It is our purpose to clarify all
the restrictions imposed by these conditions on the Maxwell tensor
$F_{ik}$. Using the plasma velocity vector $v^{i}$, we shall
introduce the vectors of the electric field $E_{i}$ and {\it
magnetic field} $H_{i}$ as observed in the frame of reference (FR)
comoving with the plasma [10]:
\begin{equation}
\label{2.9}E_{i}=v^{k} F_{ki};\hspace{1cm}  H_{i}=v^{k}
\stackrel{\ast}{F_{ki}} ,
\end{equation}
where $\stackrel{\ast}{F_{ki}}$ is a tensor dual to the Maxwell
antisymmetric tensor $F_{ki}$:
\begin{equation}
\label{2.10}\stackrel{\ast}{F_{ki}}=\frac{1}{2} \eta_{kilm}
F^{lm},
\end{equation}
and $\eta_{kilm}$ is the covarianly constant discriminant tensor
(see, e.g., [9]) satisfying the identity
\begin{equation}
\label{2.11}\eta_{iklm} \eta^{iqps}\equiv - \delta^{qps}_{klm}
\stackrel{Df}{=} -
\left\vert\begin{array}{ccc}\delta^{q}_{k}&\delta^{q}_{l}
&\delta^{q}_{m}\\ \delta^{p}_{k}&\delta^{p}_{l}&\delta^{p}_{m}\\
\delta^{s}_{k}&\delta^{s}_{l}&\delta^{s}_{m}\end{array}\right\vert.
\end{equation}
Due to (10), the vectors $E$ and $H$ are spacelike and orthogonal
to the velocity vector:
\begin{equation}
\label{2.12}(v,E)=0;\hspace{1.5cm} (v,H)=0.
\end{equation}

The relations (10) can be resolved with respect to the Maxwell
tensor [10]:
\begin{displaymath}
F_{ik} =v_{i} E_{k} - v_{k} E_{i} - \eta_{iklm} v^{l} H^{m};
\end{displaymath}
\begin{equation}
\label{2.13}\stackrel{\ast}{F_{ik}} = v_{i} H_{k} - v_{k} H_{i} +
\eta_{iklm} v^{l} E^{m} .
\end{equation}

Let us represent the EMT of the electromagnetic field as follows:
\begin{equation}
\label{2.14}\stackrel{f}{T^{i}_{k}} =\frac{1}{4\pi} (F^{i}_{~l}
F^{l}_{~k} + \frac{1}{4} \delta^{i}_{k} F^{lm} F_{lm})
\end{equation}
using three vectors $(v; E; H )$, one of which, $(v)$, is
timelike, while two others, $(E$ and $H)$, are spacelike:
\begin{displaymath}
\label{2.15}\stackrel{f}{T^{i}_{k}}  =  - \frac{1}{8\pi}
\Bigl[\delta^{i}_{k} (E^{2} + H^{2}) - 2 v^{i} v^{k} (E^{2} +
H^{2}) +
\end{displaymath}
\begin{equation}
+ 2 E^{i} E_{k} + 2 H^{i} H_{k} + 2 v^{i} \eta_{kpqs} E^{p} v^{q}
H^s + 2 v^{k} \eta^{ipqs} E_{p} v_{q} H_{s} \Bigr],
\end{equation}
where the following notations are introduced:
\begin{equation}
\label{2.16}E^{2} \stackrel{Df}{=} -(E,E);\hspace{1cm} H^{2}
\stackrel{Df}{=} -(H,H) .
\end{equation}

We shall require that the vector $v$ be an eigenvector of the EMT
(16). Contracting (16) with $v^{k}$ and taking into account the
identities (\ref{2.11}) and (\ref{2.12}), we get:
\begin{equation}
\label{2.17}\frac{1}{8\pi} [v^{i} (E^{2} + H^{2}) - 2 \eta^{ipqs}
E_{p} v_{q} H_{s}]= \varepsilon_{H} v^{i} .
\end{equation}
From (\ref{2.17}) we shall find the necessary condition for
$v^{i}$ being an eigenvector of the electromagnetic field EMT:
\begin{equation}
\label{2.18}\eta^{ipqs} E_{p} v_{q} H_{s}= \lambda
v^{i},\hspace{1cm}\forall \lambda \in{\rm R}.
\end{equation}
Contracting this relation with $v^{i}$, we get $\lambda = 0$. Thus
the necessary condition of compatibility (\ref{3}) and (\ref{4})
is:
\begin{equation}
\label{2.19}\eta^{ipqs} E_{p} v_{q} H_{s} = 0.
\end{equation}
The necessary and sufficient condition for the fulfilment of Eq.
(\ref{2.19}) is, as we know, the complanarity of the vectors $E,
H, v$, i.e.,
\begin{displaymath}
\alpha v_{i} + \beta E_{i} + \gamma H_{i} = 0.
\end{displaymath}
Contracting this relation with $v^{i}$ and taking into account
(\ref{2.6}), (\ref{2.12}) we obtain $\alpha = 0$. Thus, Eq.
(\ref{2.19}) is equivalent to the condition:
\begin{displaymath}
\beta E_{i} + \gamma H_{i} = 0.
\end{displaymath}
Since we consider {\it magnetoactive plasma}, we shall further
assume:
\begin{equation}
\label{2.20}F_{ik} F^{ik} \equiv 2(H^{2} - E^{2}) > 0.
\end{equation}
Due to (\ref{2.20}), the necessary and sufficient condition for
the fulfilment of (\ref{2.19}) is:
\begin{equation}
\label{2.21} E_{i} = \lambda H_{i};\hspace{1cm}
\forall\lambda\in{\rm R}
\end{equation}
Besides, according to (\ref{2.17}),
\begin{equation}
\label{2.22} \varepsilon_{H} = \frac{E^{2} + H^{2}}{8 \pi}.
\end{equation}
However, we have not yet extracted all the algebraic information
contained in (\ref{3}) and (\ref{4}). With due account of the
definitions of the vectors $E$ and $H$ (10), Eqs. (\ref{2.9}) can
be written as:
\begin{equation}
\label{2.23}{\cal F}_{ik} v^{k} = O,
\end{equation}
where a new antisymmetric tensor ${\cal F}_{ik}$ has been
introduced:
\begin{equation}
\label{2.24}{\cal F}_{ik} = F_{ik} - \lambda
\stackrel{\ast}{F_{ik}}.
\end{equation}
The relations (\ref{2.23}) can be regarded as a set of linear
homogeneous algebraic equations with respect to the velocity
vector $v^{i}$. The necessary and sufficient condition for a
nontrivial compatibility of these equations is:
\begin{equation}
\label{2.25}\mbox{Det}\| {\cal F}_{ik}\| = 0.
\end{equation}
As  $\| {\cal F}_{ik}\|$  is an even order antisymmetric matrix,
\begin{equation}
\label{2.26}\mbox{Det}\| {\cal F}_{ik}\| = \frac{1}{16} (\sqrt{
-g} {\cal F}^{ik} \stackrel{\ast}{\cal F}_{ik})^{2}.
\end{equation}
Therefore, the condition (\ref{2.25})) reduces to the following:
\begin{equation}
\label{2.27} \eta^{ijkl} {\cal F}_{ij} {\cal F}_{kl} = 0,
\end{equation}
-- in this case
\begin{equation}
\label{2.28}\mbox{rank} \| {\cal F}_{ik}\| = 2,
\end{equation}
i.e., the set (\ref{2.23}) admits two linearly independent
solutions for the eigenvector $v^{i}$.

Substituting into (\ref{2.27}) the expressions for $F_{ik}$ and
$\stackrel{\ast}{F}_{ik}$ from (\ref{2.13}), we get:
\begin{displaymath}
 \eta^{ijkl} {\cal F}_{ij} {\cal F}_{kl} =
4 \lambda (1 + \lambda^{2}) H^{2} = 0,
\end{displaymath}
hence follows the only possible solution under the condition
(\ref{2.20}):
\begin{displaymath}
\lambda = 0.
\end{displaymath}

Thus we come to the following rigorous conclusion. {\sl For the
EMT of elec\-tro\-mag\-ne\-tic field (15) to permit as its
eigenvector the dynamical velocity vector of the isotropic perfect
fluid under the condition (21), it is necessary and sufficient
that the electric field intensity vector in the comoving FR should
be equal to zero}:
\begin{equation}
\label{2.29} E_{i} = 0.
\end{equation}

In this case the conditions (\ref{2.23})-(\ref{2.28}) give:
\begin{equation}
\label{2.30} \stackrel{\ast}{F^{ik}} F_{ik} = 0;
\end{equation}
\begin{equation}
\label{2.31} \mbox{Det}\| F_{ik}\| = 0 \Longrightarrow \mbox{rank}
\| F_{ik}\| = 2,
\end{equation}
while the eigenvector of the fluid EMT must satisfy the set of
linear homogeneous algebraic equations:
\begin{equation}
\label{2.32} F_{ik} v^{k} = 0.
\end{equation}

Note that if Eq. (\ref{2.31}) is fulfilled, then similar
conditions for the dual Maxwell tensor are automatically fulfilled
as well:
\begin{equation}
\label{2.33} \mbox{Det}\| \stackrel{\ast}{F_{ik}} \| = 0
\Longrightarrow \mbox{rank}\| \stackrel{\ast}{F_{ik}} \| = 2.
\end{equation}

With (\ref{2.29}) write down the Maxwell tensor and the
electromagnetic field EMT (\ref{2.14}):
\begin{displaymath}
F_{ik} = - \eta_{iklm} v^{l} H^{m};
\end{displaymath}
\begin{equation}
\label{2.34} \stackrel{\ast}{F}_{ik} = v_{i} H_{k} - v_{k} H_{i};
\end{equation}
\begin{equation}
\label{2.35} \stackrel{f}{T^{i}}_{k} = \frac{1}{8\pi} (2 H^{2}
v^{i} v_{k} - 2 H^{i} H_{k} - \delta^{i}_{k} H^{2} ).
\end{equation}
Note that due to Eq. (\ref{2.34}) a more severe condition than
(\ref{2.30}) is fulfilled:
\begin{equation}
\label{2.36} \stackrel{\ast}{F_{ik}} F^{lk} = 0.
\end{equation}

The summed EMT of the magnetoactive plasma
\begin{displaymath}
T_{ik} = \stackrel{p}{T}_{ik} + \stackrel{f}{T}_{ik}
\end{displaymath}
takes the form:
\begin{equation}
\label{2.37}T_{ik} = ({\cal E} + P) v_{i} v_{k} - P g_{ik} - 2
P_{H} n_{i} n_{k},
\end{equation}
where
\begin{equation}
\label{2.38}P_{H} = \frac{H^{2}}{8 \pi};\hspace{1cm}{\cal E} =
\varepsilon + \varepsilon_{H}; \hspace{1cm} P = p + P_{H},
\end{equation}
$P$ and ${\cal E}$ being the summed pressure and energy density of
the magnetoactive plasma and
\begin{equation}
\label{2.39}n_{i} = \frac{H_{i}}{H}
\end{equation}
-- is the spacelike unit vector of magnetic field direction:
\begin{equation}
\label{2.40}(n,n) = -1,
\end{equation}
with
\begin{equation}
\label{2.41}(n,v) = 0.
\end{equation}

\section{MHD equations for plasma in a gravitational
field}
The MHD equations are to be obtained on the basis of the vanishing
divergence requirement for the summed EMT of the magnetoactive
plasma (\ref{2.37}) supplemented by the first group of the Maxwell
equations:
\begin{equation}
\label{3.1}T^{k}_{\ i,k} =0,
\end{equation}
\begin{equation}
\label{3.2}\stackrel{\ast}{F^{ik}}_{,k}=0.
\end{equation}
This set of equations with due account of the equation of state
(8), the definition of the plasma EMT (\ref{2.7}) and the
algebraic relations (\ref{2.6}), (\ref{2.12}), (\ref{2.13}),
(\ref{2.29}), (\ref{2.34}) and (\ref{2.37})  completely describes
the self-consistent motion of the magnetoactive plasma with an
embedded magnetic field in a prescribed gravitational field.
Indeed, Eqs. (\ref{3.1}), (\ref{3.2}) represent a set of 8
differential equations with respect to 10 quantities $\varepsilon,
p, H^{i},v^{i}$. However, the equation of state (\ref{2.7}), the
velocity vector normalization (\ref{2.6}), the ortogonality
condition (\ref{2.12}) and (\ref{2.29}) raise the total number of
equations up to 12. Nevertheless, it turns out that not all of
these relations are independent, as we shall see below.

To deduce the MHD equations, let us take into account the
well-known relationship (see, e.g., [11]):
\begin{equation}
\label{3.3}\stackrel{f}{T^{k}}_{i,k}= -\frac{1}{4 \pi} F_{il}
F^{kl}_{\ ,k}.
\end{equation}
Thus, Eq. (\ref{3.1}) can be presented in the form:
\begin{equation}
\label{3.4}F_{ik} \Phi^{k} = \tau_{i},
\end{equation}
where
\begin{equation}
\label{3.5}\Phi^{k} \stackrel{Df}{=} F^{kl}_{\ ,l};
\end{equation}
\begin{equation}
\label{3.6}\tau_{i} \stackrel{Df}{=} - 4 \pi
\stackrel{p}{T^{k}}_{i,k}.
\end{equation}
Eqs. (\ref{3.4}) can be regarded as a set of linear inhomogeneous
algebraic equations with respect to $\Phi^{k}$. If
$\mbox{Det}\|F_{ik}\| \not=0$, then the equations are solved in a
simple straightforward way:
\begin{equation}
\label{3.7} \Phi^{k} = \frac{{\cal A}^{k} \| F;\tau\|}{\mbox{Det}
\| F\|},
\end{equation}
where ${\cal A}^{k} \| F;\tau \|$  is a cofactor of the augmented
matrix of the set (\ref{3.4}). In particular, for the case of
vacuum ($ \tau_{i}=0$) we obtain a trivial solution: $\Phi^{k}=0$,
and the set of equations (\ref{3.1}), (\ref{3.2}) reduces to the
Maxwell equations in vacuum.

Now we pose the problem of solving the set of equations
(\ref{3.4}) with respect to $ \Phi^{k}$, i.e., the problem of
reducing Eqs. (\ref{3.1}, (\ref{3.2}) to the form of the Maxwell
equations {\sl with minimal requirements upon the electomagnetic
field invariants}:
\begin{equation}
\label{3.8} F^{ik} \stackrel{\ast}{F}_{ik}=0,
\end{equation}
\begin{equation}
\label{3.9} F_{ik} F^{ik} > 0.
\end{equation}
The existence of a {\it positive} invariant (\ref{3.9}) means that
we can choose a local FR where the electric field is absent [11].

Provided (50) is fulfilled, as before (see (\ref{2.31})),
\begin{equation}
\label{3.10} \mbox{Det}\| F \| = 0 \Longrightarrow
\end{equation}
\begin{equation}
\label{3.11} \mbox{rank}\| F \| =2.
\end{equation}
Thus for the consistency of the algebraic set of equations
(\ref{3.4}) under the condition (\ref{3.8}) it is {\it necessary
and sufficient} that:
\begin{equation}
\label{3.12} \mbox{rank}\| F;\tau \|=2.
\end{equation}
Calculating all the 3rd order minors of the augmented matrix  $ \|
F;\tau \|$ with (\ref{3.8}), we obtain a condition equivalent to
(\ref{3.12}):
\begin{equation}
\label{3.13} \stackrel{\ast}{ F}_{ik} \tau_{k} =0
\Longleftrightarrow \mbox{rank}\| F;\tau \| =2.
\end{equation}

To solve the set of equations (\ref{3.4}), consider the
eigenvectors of the matrix  $\|F_{ik}\|$:
\begin{equation}
\label{3.14} F_{ik} u^{k} = \lambda u_{i}.
\end{equation}
Due to the antisymmetry of $F_{ik}$, it follows from (\ref{3.14})
that either $\lambda =0$, or $u$ is a null vector. It can be
demonstrated that provided Eqs. (\ref{3.8}) and (\ref{3.9}) are
valid, $u$ cannot be a null vector. Thus, the {\it Maxwell tensor
admits only nonnull eigenvectors with zero eigenvalues}:
\begin{equation}
\label{3.15} F_{ik} u^{k} = 0.
\end{equation}
By (\ref{3.8}),(\ref{3.10}) and (\ref{3.11}), this eigenvalue is
doubly degenerate, and thus, according to a well-known algebraic
theorem, two lineary independent eigenvectors cor\-res\-pond to
it: $\stackunder{(1)}{u}$ and $\stackunder{(2)}{u}$.

Under the condition (\ref{3.9}) we can always choose a local FR
where $F_{i4}=0$. In this FR there always exists an eigenvector of
the Maxwell tensor of the form $u^{k}=\delta^{k}_{4}$. Therefore,
one of the eigenvectors of the matrix  $\| F \|$, e.g.,
$\stackunder{(1)}{u}$, is timelike and the second one,
$\stackunder{(2)}{u}$, is spacelike. Using the standard
orthogonalization process, we normalize them as follows:
\begin{equation}
\label{3.16}
(\stackunder{(1)}{u},\stackunder{(1)}{u})=1;\hspace{1cm}
(\stackunder{(2)}{u},\stackunder{(2)}{u})= -1;\hspace{1cm}
(\stackunder{(1)}{u},\stackunder{(2)}{u})= 0.
\end{equation}
Then the general solution of (\ref{3.15}) can be written in the
form
\begin{equation}
\label{3.17}u^{k} = \alpha \stackunder{(1)}{u^{k}} + \beta
\stackunder{(2)}{u^{k}},
\end{equation}
where $\alpha$ and $\beta$ are arbitary scalars.

Let us now investigate the relations (\ref{3.13}), which can be
regarded as algebraic equations with respect to $\tau$. Since
\begin{equation}
\label{3.18}-F_{ik} F^{ik} = \stackrel{\ast}{F}_{ik}
\stackrel{\ast}{F^{ik}} <0,
\end{equation}
and Eq. (\ref{3.8}) is invariant under the substitution $ F
\leftrightarrow  \stackrel{\ast}{F} $, as well as the expression
for $ \mbox{Det}\|F\|$ (\ref{2.26}), we conclude that the dual
matrix $\|\stackrel{\ast}{F_{ik}}\|$  also admits two and only two
linearly independent spacelike eigenvectors $\stackunder{(1)}{w}$
and $\stackunder{(2)}{w}$, which correspond to a zero eigenvalue.
It is not difficult to verify (e.g., turning to a FR where $
F_{\alpha 4}=0 $) that the rank of the unified matrix
$\|F,\stackrel{\ast}{F}\|$ under the condition (\ref{3.8}) is
equal to 4. Hence the eigenvectors of the matrices $ \|F\| $ and $
\| \stackrel{\ast}{F} \| $  are linearly independent and we can
choose the following normalization for them:
\begin{displaymath}
(\stackunder{(1)}{w},\stackunder{(1)}{w})= - 1;\hspace{1cm}
(\stackunder{(2)}{w},\stackunder{(2)}{w})= - 1;\hspace{1cm}
\end{displaymath}
\begin{equation}
\label{3.19}(\stackunder{(1)}{w},\stackunder{(2)}{w})=0;\hspace{1cm}
(\stackunder{(\alpha)}{w},\stackunder{(\beta)}{u}) =
0,\hspace{1cm} (\alpha,\beta = 1,2).
\end{equation}
Thus, the general solution to Eq. (\ref{3.13}) is:
\begin{equation}
\label{3.20} \tau_{i}= \lambda \stackunder{(1)}{w_{i}} + \mu
\stackunder{(2)}{w_{i}},
\end{equation}
where $\lambda$ and $\mu$ are arbitary scalars and due to
(\ref{3.19}):
\begin{equation}
\label{3.21}(\tau,\stackunder{(\alpha)}{u})=0.
\end{equation}

By (\ref{3.11}) and (\ref{3.13}), the Maxwell tensor and its dual
can be represented in terms of the eigenvectors of the matrices
$\| F \|$  and  $\| \stackrel{\ast}{F} \|$:
\begin{equation}
\label{3.22}F_{ik}= - \sigma \eta_{iklm} \stackunder{(1)}{u}^{l}
\stackunder{(2)}{u}^{m};
\end{equation}
\begin{equation}
\label{3.23} \stackrel{\ast}{F}_{ik} = \varrho \eta_{iklm}
\stackunder{(1)}{w}^{l} \stackunder{(2)}{w}^{m},
\end{equation}
where $\sigma$ and $\varrho$ are certain scalars. Contracting
these relations with the discriminant tensor, we obtain the dual
relations:
\begin{equation}
\label{3.24} \stackrel{\ast}{F^{ik}}= \sigma
(\stackunder{(1)}{u}^{i} \stackunder{(2)}{u}^{k} -
\stackunder{(2)}{u}^{i} \stackunder{(1)}{u}^{k});
\end{equation}
\begin{equation}
\label{3.25} F^{ik}= - \varrho (\stackunder{(1)}{w}^{i}
\stackunder{(2)}{w}^{k} - \stackunder{(2)}{w}^{i}
\stackunder{(1)}{w}^{k}).
\end{equation}

In particular, the following relation stems from (\ref{3.22}) and
(\ref{3.24}):
\begin{equation}
\label{3.26} \stackrel{\ast}{F_{ik}} F^{kl}=0.
\end{equation}

Using the Maxwell tensor representation (\ref{3.22}) and the
orthonormality relations (\ref{3.16}), we get the useful formula:
\begin{equation}
\label{3.27} F_{ik} F^{il}= \sigma^{2} (- \stackunder{(1)}{u_{k}}
\stackunder{(1)}{u^{l}} + \stackunder{(2)}{u_{k}}
\stackunder{(2)}{u^{l}} + \delta^{l}_{k}).
\end{equation}
Contracting (\ref{3.27}), we get:
\begin{equation}
\label{3.28} \frac{1}{2} F_{ik} F^{ik} = \sigma^{2} >0.
\end{equation}
Contracting Eqs (\ref{3.4}) with $F^{il}$ and taking into account
(\ref{3.27}) and (\ref{3.28}), we obtain an equation equivalent to
Eq. (\ref{3.4}):
\begin{equation}
\label{3.29} \sigma^{2} [\Phi^{l} -
 \stackunder{(1)}{u^{l}}(\stackunder{(1)}{u},\Phi) +
\stackunder{(2)}{u^{l}}(\stackunder{(2)}{u},\Phi)] = F^{il}
\tau_{i}.
\end{equation}
A special solution to Eq. (\ref{3.29}) is:
\begin{equation}
\label{3.30} \Phi^{i}_{(1)}= \frac{1}{\sigma^{2}} F^{il} \tau_{i}.
\end{equation}
Therefore the general solution of Eqs. (\ref{3.4}) can be
presented in the form:
\begin{equation}
\label{3.31} F^{ik}_{, k} \equiv \Phi^{i} = \frac{8 \pi F^{ik}
\stackrel{p}{T^{j}_{k,j}}}{F^{lm} F_{lm}}   + \alpha
\stackunder{(1)}{u^{i}}+ \beta \stackunder{(2)}{u^{i}}.
\end{equation}

This exhausts the problem of reducing the set of equations
(\ref{3.1}),(\ref{3.2})  to the standard Maxwell form.

If the external currents are absent $(\alpha=\beta=0)$, then Eq.
(\ref{3.31}) reduces to the form of the second group Maxwell
equations:
\begin{equation}
\label{3.32} F^{ik}_{, k}= - 4 \pi J^{i}_{dr},
\end{equation}
where:
\begin{equation}
\label{3.33} J^{i}_{dr} = - \frac{2 F^{ik}
\stackrel{p}{T^{l}}_{k,l} }{F_{jm} F^{jm}} -
\end{equation}
is the {\em drift current}, which by (\ref{3.26}) satisfies the
relation:
\begin{equation}
\label{3.34} \stackrel{\ast}{F}_{ij} J^{i}_{\mbox{\em dr}}=0,
\end{equation}
Hence, due to (\ref{3.13}) and (\ref{3.21}:
\begin{equation}
\label{3.35} (J_{\mbox{\em dr}}, \stackunder{(\alpha)}{u}) =0,
\end{equation}
\begin{equation}
\label{3.36} (J_{dr}, J_{dr}) <0 -
\end{equation}
i.e., {\em the drift current is spacelike}.

Calculating the covariant divergence $\nabla_{l}$ of Eq.
(\ref{3.26}) with the aid of the first group Maxwell equations
(\ref{3.2}), we get the differential implication:
\begin{equation}
\label{3.37} \stackrel{\ast}{F^{kl}} F_{kl,i} = F^{kl}
\stackrel{\ast}{F}_{kl,i} =0.
\end{equation}
Calculating the covariant divergence  $\nabla^{i}$ of (\ref{3.26})
using the second group Maxwell equations (\ref{3.32}) and the
relation (\ref{3.34}), we get one more differential implication:
\begin{equation}
\label{3.38} F^{kl} \stackrel{\ast}{F}_{i[k,l]} =0.
\end{equation}

Setting in Eqs. (\ref{3.15}),(\ref{3.16}),
(\ref{3.19}),(\ref{3.22})-(\ref{3.25}) $ \stackunder{(1)}{u}^{i}
=v^{i}$, $ \stackunder{(2)}{u}^{i}=n^{i} \equiv (H^{i}/H)$, we
find a complete coincidence of the above formulas with the
relevant expressions from the previous section. Thus:
\begin{equation}
\label{3.39} \sigma^{2} = \frac{1}{2} F_{lm} F^{lm}= - (H,H)
\equiv H^{2}.
\end{equation}

Thus due to (\ref{3.21}) the following differential relations
should be valid:
\begin{equation}
\label{3.40} v^{i} \stackrel{p}{T^{k}}_{i,k} =0,
\end{equation}
\begin{equation}
\label{3.41} H^{i} \stackrel{p}{T^{k}}_{i,k} =0.
\end{equation}
Substituting to (\ref{3.40}) and (\ref{3.41}) the expression for $
\stackrel{p}{T}_{ik}$ (\ref{2.5}) and taking into account
(\ref{2.6}) and (\ref{2.12}), we obtain:
\begin{equation}
\label{3.42} v^{k}_{,k} = - \frac{\varepsilon_{,k} v^{k}}
{\varepsilon + p};
\end{equation}
\begin{equation}
\label{3.43} v_{i,k} H^{i} v^{k} = \frac{p_{,i} H^{i}}
{\varepsilon + p}.
\end{equation}

Calculating the drift current (\ref{3.33}) with Eqs.
(\ref{2.5}),(\ref{2.32}) and (\ref{3.39})), we get:
\begin{equation}
\label{3.44} J^{i}_{dr}= -2 \frac{F^{ik} [v_{k,l}v^{l}(\varepsilon
+ p) - p_{,k}]}{F_{lm} F^{lm}}.
\end{equation}

From (\ref{2.34}) we obtain the useful relation (for $H \not=0$):
\begin{equation}
\label{3.45} v^{i} = \frac{\stackrel{\ast}{F^{ki}} H_{k}}{H^{2}}.
\end{equation}
By (35) and the definition of the vector $H^{i}$ (10) the
orthonormality relations (\ref{2.12}) are fulfilled identically.

Note that due to (\ref{3.26}) the solution to Eq. (\ref{2.34}) is
\begin{equation}
\label{3.46} v^{i}= \stackrel{\ast}{F^{ki}} S_{k},
\end{equation}
where  $S_{k}$ is an arbitary spacelike vector, satisfying the
only condition:
\begin{equation}
\label{3.47} \stackrel{\ast}{F^{ki}} \stackrel{\ast}{F^{l}_{\ i}}
S_{k} S_{l} =1.
\end{equation}

Now let us turn to the first group of the Maxwell equations
(\ref{3.2}). Using the representation (\ref{2.34}) for the Maxwell
dual tensor, we obtain for Eq. (\ref{3.2}):
\begin{equation}
\label{3.48} v^{i} H^{k}_{,k} + v^{i}_{,k} H^{k} - v^{k}_{,k}
H^{i} - v^{k} H^{i}_{,k} =0.
\end{equation}
Consecutively contracting Eqs. (\ref{3.48}) with $v^{i}$ and
$H^{i}$ and using (\ref{2.6}) and (\ref{2.12}), we obtain:
\begin{equation}
\label{3.49} -v_{i,k} H^{i} v^{k} =H_{i,k} v^{i} v^{k}=H^{k}_{,k};
\end{equation}
\begin{equation}
\label{3.50} H_{i,k} v^{i} H^{k}= -v_{i,k} H^{i} H^{k}=H(H
v^{k})_{,k}.
\end{equation}
Since the rank of the matrix $\| \stackrel{\ast}{F_{ik}}\|$ equals
2, the relations (\ref{3.48}) -- (\ref{3.50}) are equivalent to
the first group of the Maxwell equations (\ref{3.2}). By
(\ref{3.49}), Eq. (\ref{3.43}) reduces to a form similar to
(\ref{3.42}):
\begin{equation}
\label{3.51} H^{k}_{,k}= - \frac{p_{,k} H^{k}}{\varepsilon + p}.
\end{equation}

\section{Solution of magnetohydrodynamic equations\newline
against the background of a plane gravitational wave}
\subsection{Initial conditions and symmetry of the problem}
The metric of a PGW with the polarization ${\bf e}_{+}$ is
described by the expression [12]:
\begin{equation}
\label{4.1} d s^{2}=2 du dv - L^{2}[e^{2 \beta}(dx^{2})^{2} +
e^{-2\beta}( dx^{3})^{2}],
\end{equation}
where $\beta(u)$ is an arbitrary function (the PGW amplitude); the
function $L(u)$ (the PGW background factor) obeys an ordinary
second-order differential equation; $ u= \frac{1}{\sqrt{2}}(t -
x^{1})$ is the retarted time and $v= \frac{1}{\sqrt{2}}(t +
x^{1})$ is the advanced time. The absolute future is represented
by the region ${\it T^{+}}:\lbrace u>0;v>0\rbrace$, the absolute
past by ${\it T^{-}}:\lbrace u<0;v<0\rbrace$. The metric
(\ref{4.1}) admits the group of motions $ {\cal G}_{5}$,
associated with {\em three linearly independent Killing vectors}:
\begin{equation}
\label{4.2} \stackunder{(1)}{\xi^{i}} =\delta^{i}_{v}; \hspace{1
cm} \stackunder{(2)}{\xi^{i}} = \delta^{i}_{2}; \hspace{1 cm}
\stackunder{(3)}{\xi^{i}} = \delta^{i}_{3}.
\end{equation}

Let there be no GW at $u \leq 0$, i.e.,
\begin{equation}
\label{4.3} \beta(u)_{\mid u \leq 0}=0; \hspace{1 cm} L(u)_{\mid u
\leq 0}=1,
\end{equation}
the plasma is homogeneous and at rest:
\begin{displaymath}
v^{v}_{\mid u \leq 0}= v^{u}_{\mid u \leq 0}= \frac{1}{\sqrt{2}};
\hspace{1 cm} v^{2}=v^{3}=0;
\end{displaymath}
\begin{equation}
\label{4.4} \varepsilon_{\mid u \leq 0}=\varepsilon_{0};
\hspace{1cm} p_{\mid u \leq 0}=p_{0},
\end{equation}
and a homogeneous magnetic field vector belongs to the plane
$\lbrace x^{1},x^{2} \rbrace $:
\begin{displaymath}
H_{1 \mid u \leq 0}=H_{0} \cos\Omega ; \hspace{1 cm} H_{2 \mid u
\leq 0}=H_{0} \sin\Omega ;
\end{displaymath}
\begin{equation}
\label{4.5} H_{3 \mid u \leq 0}=0; \hspace{1 cm} E_{i \mid u \leq
0}=0,
\end{equation}
where $\Omega$ is the angle between the axis $0x^{1}$ (the PGW
propagation direction) and the direction of the magnetic field
${\bf H}$. The conditions (\ref{4.5}) agree with the vector
potential:
\begin{displaymath}
A_{v}=A_{u}=A_{2}=0;
\end{displaymath}
\begin{equation}
\label{4.6} A_{3}=H_{0} (x^{1} \sin\Omega - x^{2} \cos\Omega);
\hspace{1.5 cm} (u \leq 0).
\end{equation}
In [7] it is demonstrated that the solution to the MHD equations
in the metric (\ref{4.1}) with the initial conditions (\ref{4.3})
-- (\ref{4.6}) with $\varepsilon \not\equiv 0 $ is {\em strictly
stationary}, i.e., all observed quantities are functions of solely
the retarded time $u$. Therefore, we shall immediately require
that the solution of our problem {\em inherit the symmetry of the
metric} (\ref{4.1}):
(\ref{4.1}):
\begin{equation}
\label{4.7} \stackunder{\xi_{(\alpha)}}{\cal L} {\bf P} =0;
\hspace{1,5 cm} (\alpha =1,2,3)
\end{equation}
for all observed quantities ${\bf P}$ ($ \stackunder{\xi}{\cal L}$
is a Lie derivative in the direction $ \xi $):
\begin{equation}
\label{4.8} p=p(u); \hspace{1 cm} \varepsilon=\varepsilon(u);
\hspace{1 cm}v^{i}=v^{i}(u);
\end{equation}
\begin{equation}
\label{4.9} F_{ik}=F_{ik}(u).
\end{equation}

With (\ref{4.9}) we obtain the following results from the first
group of the Maxwell equations:
\begin{equation}
\label{4.10} L^{2} \stackrel{\ast}{F^{u\alpha}} = {\cal
C}_{(\alpha)} \hspace{1 cm}(=Const); \hspace{1 cm} \alpha =
\lbrace v,2,3 \rbrace.
\end{equation}
Thus, using the initial conditions (\ref{4.5}) - (\ref{4.6}), we
find:
\begin{displaymath}
L^{2} \stackrel{\ast}{F^{uv}}= - F_{23}= H_{0} \cos\Omega ;
\end{displaymath}
\begin{displaymath}
L^{2} \stackrel{\ast}{F^{u2}}= F_{v3}= \frac{1}{\sqrt2}
H_{0}\sin\Omega ;
\end{displaymath}
\begin{equation}
\label{4.11} L^{2} \stackrel{\ast}{F^{u3}}= -F_{v2}= 0.
\end{equation}

The condition (\ref{3.8}) ($ \stackrel{\ast}{F_{ik}} F^{ik}=0 $)
with (\ref{4.11}) reduces to:
\begin{equation}
\label{4.12} L^{2} \stackrel{\ast}{F^{v3}}= F_{u2}= {\sqrt 2}
F_{uv} \cot\Omega ,
\end{equation}
while the MFEP conditions (\ref{2.29}) with (\ref{4.11}) and
(\ref{4.12}) yield:
\begin{equation}
\label{4.13} v^{3} = \frac{{\sqrt 2} F_{uv}}{H_{0} \sin\Omega}
v_{v};
\end{equation}
\begin{equation}
\label{4.14} \frac{1}{\sqrt{ 2}} H_{0} v_{u} \sin\Omega  + F_{u3}
v_{v} - H_{0} v^{2} \cos\Omega =0.
\end{equation}

\subsection{Conservation laws and the vector potential}
A consequence of the second group of the Maxwell equations
(\ref{3.32}) is, as we know, the current conservation law, which,
in view of (\ref{4.7}) and the initial conditions (\ref{4.4}) --
(\ref{4.5}), takes the form:
\begin{equation}
\label{4.15} J^{u}_{ dr} =0.
\end{equation}
Calculating $ J^{u}_{dr} $ using (\ref{3.44}), (\ref{4.11}) and
(\ref{4.13}), we reduce (\ref{4.15}) to the form:
\begin{equation}
\label{4.16} v_{v} (L^{2} e^{-2 \beta} F_{uv})'=0,
\end{equation}
where a prime is a derivative with respect to $u$. Thus, due to
the initial conditions ($ F_{uv}(u)_{\vert u\leq 0}=0 $) and the
the velocity vector timelikeness requirement ($ v_{v} \not\equiv 0
$) we draw the conclusion that the current conservation law
(\ref{4.15}) is equivalent to the requirement:
\begin{equation}
\label{4.17} F_{uv} =0.
\end{equation}
But then by (\ref{4.12}) and (\ref{4.13})
\begin{equation}
\label{4.18} \stackrel{\ast}{F^{v3}} = F_{u2} =0;
\end{equation}
\begin{equation}
\label{4.19} v^{3} =0.
\end{equation}
As we know, the first group of the Maxwell equations (\ref{3.2})
is equivalent to the existence condition of a vector potential $
A_{i} $:
\begin{equation}
\label{4.20} F_{ik} = \partial_{i} A_{k} - \partial_{k} A_{i}.
\end{equation}
Thus, we can write for the zero component of the Maxwell tensor
(\ref{4.11}), (\ref{4.17}), (\ref{4.18}):
\begin{equation}
\label{4.21}
\partial_{\sigma}A_{\gamma}-\partial_{\gamma}A_{\sigma} =
0;\hspace{1 cm} (\lbrace \gamma,\sigma \rbrace =\lbrace u,v,2
\rbrace).
\end{equation}
As known, a unique solution to Eqs. (\ref{4.21}) on the
3-dimensional hypersurface $ \Sigma_{3}:\lbrace x^{3}=\mbox{Const}
\rbrace $ is a gradient vector:
\begin{equation}
\label{4.22} A_{\sigma} = \partial_{\sigma}\Phi, \hspace{1 cm}
(\sigma=u,v,2),
\end{equation}
where $ \Phi=\Phi(u,v,x^{2},x^{3}) $  is an arbitrary scalar
function.

The nonzero components of the Maxwell tensor, $F_{\sigma 3}$, can
be represented, by (\ref{4.22}), as:
\begin{equation}
\label{4.23} F_{\sigma 3} = \partial_{\sigma} {\tilde
A}_{3},\hspace{1 cm} (\sigma=\lbrace v,u,2 \rbrace)
\end{equation}
where
\begin{equation}
\label{4.24} {\tilde A}_{3} \stackrel{Df}{=} A_{3} -
\partial_{3}\Phi
\end{equation}
is a gradient-renormalized vector potential. Calculating the
nonzero components of the Maxwell tensor $F_{\sigma 3}$ using
(\ref{4.23}) - (\ref{4.24}) and taking into account the relations
(\ref{4.9}) and (\ref{4.11}), we finally find:
\begin{equation}
\label{4.25} {\tilde A}_{3} = - H_{0} x^{2} \cos\Omega +
\frac{1}{\sqrt{2}} H_{0} [v - \psi(u)] \sin\Omega,
\end{equation}
where $\psi(u)$ is an arbitrary differentiable function satisfying
the initial condition
\begin{equation}
\label{4.26} \psi_{|u\leq 0} = u.
\end{equation}
It should be noted that the only $\psi$-dependent nonvanishing
component of the Maxwell tensor is
\begin{equation}
\label{4.27} F_{u3} = - \frac{1}{\sqrt{2}}H_{0}\psi'\sin\Omega,
\end{equation}
and the MFEP condition (\ref{4.14}) takes the form:
\begin{equation}
\label{4.28} \frac{1}{\sqrt{2}}(v_{v}\psi' - v_{u})\sin\Omega +
v^{2} \cos\Omega = 0.
\end{equation}

Calculating the other components of the drift current using
((\ref{4.11}),(\ref{4.17}) and (\ref{4.18}), we find:
\begin{equation}
\label{4.29} J^{v} = J^{2} = 0,
\end{equation}
and for the only nontrivial component $J^{3}$ we obtain an
expression coinciding with the one found in [6] in the case of
rigorously transverse PGW propagation ($\cos\Omega =0$). However,
for other values of $\Omega$ the expression for the drift current
obtained in [6], as well as that for the drift velocity (cf.
(\ref{4.28})), is erroneous. However, we shall not integrate the
Maxwell equations with the drift current, since it is a
consequence of the conservation laws (\ref{3.1}) and the MFEP
conditions: it is much simpler in our case to integrate the
conservation laws directly.

\subsection{Integrals of motion}
Due to the conservation laws for the complete plasma EMT
(\ref{3.1}) and the presence of three linearly independent Killing
vectors (\ref{4.2}) there are 3 integrals of Eqs. (\ref{3.1}):
\begin{equation}
\label{4.30} L^{2} \stackunder{(\alpha)}{\xi^{k}} T^{u}_{k} =
\stackunder{(\alpha)}{C};\hspace{1 cm}  (\alpha =1,2,3),
\end{equation}
where $T^{u}_{k}$ is described by Eqs. (\ref{2.35}) and
(\ref{2.37}):
\begin{equation}
\label{4.31} T^{u}_{k}=\left(\varepsilon +p+
\frac{H^{2}}{4\pi}\right)v_{v}v_{k} -
\left(p+\frac{H^{2}}{8\pi}\right)\delta^{u}_{k} - \frac{H_{v}
H_{k}}{4\pi}.
\end{equation}
Calculating the magnetic field vector $H^{i}$ and the scalar H2
according to (\ref{2.9}) and (\ref{3.39}), we find:
\begin{displaymath}
H_{v}=-H_{0} L^{-2} (v_{v} \cos\Omega + \frac{1}{\sqrt{2}} v_{2}
\sin\Omega);
\end{displaymath}
\begin{displaymath}
H_{u}=H_{0} L^{-2} (v_{u}\cos\Omega - \frac{1}{\sqrt{2}} \psi'
v_{2} \sin\Omega);
\end{displaymath}
\begin{equation}
\label{4.32} H_{2}= -\frac{1}{\sqrt{2}} H_{0} e^{2\beta}\sin\Omega
(v_{u}+v_{v} \psi'); \hspace{1.5 cm} H_{3}=0;
\end{equation}
\begin{equation}
\label{4.33} H^{2} \stackrel{Df}{=} \frac{1}{2} F_{ik} F^{ik} =
H^{2}_{0}\left(\frac{\cos^{2}\Omega}{L^{4}}+
\frac{\sin^{2}\Omega}{L^{2}} \psi' e^{2\beta}\right).
\end{equation}
It is not difficult to verify that according to (\ref{4.32}) the
orthogonality relation (\ref{2.12}) ($(v,H)=0$) is satisfied
identically. The velocity vector normalization condition
(\ref{2.6}) with (\ref{4.28}) and (\ref{4.32}) can be written in
the form:
\begin{equation}
\label{4.34} (v_{v}\cos\Omega+ \frac{1}{\sqrt{2}}v_{2}
\sin\Omega)^{2}= \frac{H^{2}}{H^{2}_{0}}v^{2}_{v} L^{4}-
\frac{\sin^{2}\Omega}{2}L^{2} e^{2\beta}.
\end{equation}
One of the integrals (\ref{4.30}), $\stackunder{(3)}{C}$, proves
to be trivial, while the remaining two yield in view of
(\ref{4.28}), (\ref{4.33}) and (\ref{4.34}):
(\ref{4.28}), (\ref{4.33}), (\ref{4.34}):
\begin{equation}
\label{4.35} (\varepsilon+p) L^{2} v^{2}_{v} +
\frac{H^{2}_{0}\sin^{2}\Omega}{8\pi}
e^{2\beta}=\stackunder{(1)}{C} =\frac{1}{2}
\left(\varepsilon_{0}+p_{0} + \frac{H^{2}_{0}
\sin^{2}\Omega}{4\pi}\right);
\end{equation}
\begin{equation}
\label{4.36} (\varepsilon+p) L^{2} v_{v} v_{2} - \frac{H^{2}_{0}
\cos\Omega \sin\Omega}{4\sqrt{2}\pi}e^{2\beta} =
\stackunder{(2)}{C}= -\frac{H^{2}_{0}\cos\Omega
\sin\Omega}{4\sqrt{2}\pi}.
\end{equation}
Thus:
\begin{equation}
\label{4.37} v^{2}_{v}=\frac{\varepsilon_{0}+p_{0}}{2
L^{2}(\varepsilon+p)} \Delta(u);
\end{equation}
\begin{equation}
\label{4.38} \frac{v_{2}}{v_{v}}=
\sqrt{2}(\Delta^{-1}-1)\cot\Omega,
\end{equation}
where:
\begin{equation}
\label{4.39} \Delta(u) \stackrel{Df}{=} 1 - \alpha^{2}(e^{2\beta}
- 1)
\end{equation}
and a dimensionless parameter $\alpha$ has been introduced:
%
\begin{equation}
\label{4.40} \alpha^{2}=
 \frac{H^{2}_{0}\sin^{2}\Omega}{4\pi(\varepsilon_{0} + p_{0})}.
\end{equation}

Let us now turn to the consequences of the Maxwell equations.
Integrating (\ref{3.43}), we find one more integral:
\begin{equation}
\label{4.41} \sqrt{2} L^{2}|v_{v}|=
\exp\left[-\int\limits_{\varepsilon_{0}}^{\varepsilon}
\frac{d\varepsilon}{\varepsilon+p(\varepsilon)}\right].
\end{equation}
Thus, if an equation of state is specified, $p=p(\varepsilon)$,
(\ref{2.7}), then using (\ref{4.37}), (\ref{4.38}) and
(\ref{4.41}), the functions $v_{v}(u),  v_{2}(u)$,
$\varepsilon(u)$ and $p(u)$ are determined. However, to be able to
determine $v_{u}(u)$, it is necessary to find the function
$\psi(u)$, for which yet one more integral is required. It is
precisely the normalization relationship in the form (\ref{4.34})
that is the integral in question. Multiplying (\ref{4.35}) by
$\cos\Omega$ and (\ref{4.36}) by $\frac{1}{\sqrt{2}}\sin\Omega$
and adding up the values thus obtained, we get\footnote{It is
easily demonstrated that this integral is directly obtainable from
(\ref{3.51}).}:
\begin{equation}
\label{4.42} L^{2} v_{v}(v_{v}\cos\Omega +
\frac{1}{\sqrt{2}}v_{2}\sin\Omega)=
\frac{\varepsilon_{0}+p_{0}}{2(\varepsilon+p)} \cos\Omega.
\end{equation}
Squaring both parts of (\ref{4.42}) and taking into account the
normalization condition in the form (\ref{4.34}) along with
(\ref{4.37}), we find:
\begin{equation}
\label{4.43} H^{2}= \frac{H^{2}_{0}}{\Delta} \left(
\frac{\cos^{2}\Omega}{L^{4}\Delta} + \frac{\varepsilon
+p}{\varepsilon_{0}+p_{0}} e^{2\beta} \sin^{2}\Omega \right).
\end{equation}
Comparing (\ref{4.43}) and (\ref{4.33}), we finally obtain:
\begin{equation}
\label{4.44} \psi'= \frac{L^{2}(\varepsilon
+p)}{\Delta(\varepsilon_{0}+p_{0})} + \left(\frac{1}{\Delta^{2}}
-1 \right)e^{-2\beta} \frac{\cot^{2}\Omega}{L^{2}}.
\end{equation}
Finally, using (\ref{4.28}),(\ref{4.38}) and (\ref{4.44}), we find
an expression for $v_{u}$:
\begin{equation}
\label{4.45} \frac{v_{u}}{v_{v}}= \frac{L^{2}(\varepsilon
+p)}{\Delta(\varepsilon_{0}+p_{0})} + \left(\frac{1}{\Delta} -
1\right)^{2} e^{-2\beta} \frac{\cot^{2}\Omega}{L^{2}}.
\end{equation}

Thus, with a specifeied equation of state (\ref{2.7}), all unknown
functions are found, the solutions to Eqs.
((\ref{4.37}),(\ref{4.38}),(\ref{4.41})--(\ref{4.45})
automatically satisfying the initial conditions
(\ref{4.4}),(\ref{4.5}) (see also (\ref{4.26})). {\em Thereby we
have found by quadratures an exact solution to the self-consistent
problem of motion of a magnetoactive plasma against the background
of a PGW}.

\section{A study of the solution}
Squaring both parts of Eq. (\ref{4.41}) and using, in the lefthand
side of the equation thus obtained, the expression for $v^{2}_{v}$
from (\ref{4.37}), we represent (\ref{4.41}) after certain obvious
transformations in the form:
\begin{equation}
\label{5.1} \Lambda(u)=e^{-\Je},
\end{equation}
where $\Lambda \stackrel{Df}{=} L^{2}(u) \Delta(u)$,
\begin{equation}
\label{5.2} \Je = \Jeoe \drob d \varepsilon.
\end{equation}
With a specified equation of state (\ref{2.7}), Eq. (\ref{5.1})
completely determines the functions $\varepsilon(u)$ and $p(u)$
and thus explicitly determines the solution to the problem posed.
Let us investigate these functions, making the most general
assumption on the equation of state:
\begin{equation}
\label{5.3} p(\varepsilon) < \varepsilon .
\end{equation}
Then
\begin{equation}
\label{5.4} p'_{\varepsilon} < 1 ,
\end{equation}
\begin{displaymath}
\drob > 0
\end{displaymath}
and thus
\begin{equation}
\label{5.5} \mbox{sgn}[\Je] = \mbox{sgn}(\varepsilon -
\varepsilon_{0})
\end{equation}
It can be seen from (\ref{5.1}) that $\varepsilon$ depends on the
retarded time only through the function $\Lambda(u)$:
\begin{equation}
\label{5.6} \varepsilon =\varepsilon(\Lambda(u)).
\end{equation}

\subsection{Singularity investigation}
Let us ivestigate the dependence $\varepsilon(\Lambda)$. It
follows from (\ref{4.37}) that the solution is specified in the
interval $\Lambda \in [0,+ \infty)$. As $\Lambda \rightarrow +0$,
according to (\ref{5.1}) $\Je \rightarrow + \infty $, which by
(\ref{5.5}) is possible only with $\varepsilon \rightarrow +
\infty $. Thus we can draw a general conclusion that {\em the
solution of the MHD equations against the background of a PGW
contains a physical singularity on the hypersurfaces}
$u=u_{\ast}$:
\begin{equation}
\label{5.7} \Lambda(u_{\ast}) = 0.
\end{equation}
By definition of the functions  $\Lambda(u)$ and $\Delta(u)$,
there can be two types of such hypersurfaces:
\begin{equation}
\label{5.8} \begin{array}{cc} A). & L^{2}(u) = 0; \\
B). & 1 - \alpha^{2}(e^{2\beta(u)} - 1) = 0. \end{array}
\end{equation}

The first type of singularity is well-known: it is connected with
a coordinate singularity of the metric (\ref{4.1}) and always
arises in plasma (see, for instance, [1]). The second type is new
and is not connected with a coordinate singularity of the PGW
metric (\ref{4.1}) -— this is a purely physical singularity [13].
By (\ref{5.8}), the conditions for the formation of a second-type
singularity are:
\begin{equation}
\label{5.9} \beta(u) > 0;
\end{equation}
\begin{equation}
\label{5.10} \alpha^{2} > 1.
\end{equation}
It is well-known (see, e.g., [12]) that the values of $\beta(u)>
0$ correspond to a compression phase of the geodesic tube along
the Ox2 axis, while $0x^{3}$, $ \beta< 0$ correspond to an
expansion phase. But the condition (\ref{5.10}), according to
(\ref{4.40}), means that in the initial state $
\varepsilon_{H}\sin^{2}\Omega > \frac{1}{2}(\eopo)$, i.e., the
plasma is highly magnetized. It is an extremely important fact
that the $B$ - type singular state is possible even in a weak PGW
$(|\beta|\ll 1)$ provided the plasma is highly magnetized
$(\alpha^{2} \gg 1)$; in this case, according to (\ref{5.8}), the
singular state occurs on the hypersurfaces $ u=u_{\ast}$:
\begin{equation}
\label{5.11} \beta(u_{\ast}) =  \frac{1}{2 \alpha^{2}}.
\end{equation}
Further, according to (\ref{5.1}) and (\ref{5.2}), at
$\Lambda(u)=1$: $\varepsilon=\varepsilon_{0}, p=p_{0} $ and thus
by (\ref{4.37})--(\ref{4.45}) the initial conditions are restored
on the hypersurfaces $ \Lambda(u) = 1 $. As $ \Lambda(u)
\rightarrow + \infty $, by (\ref{5.1}) $ \Je \rightarrow - \infty
$, which is possible by (\ref{5.5}) only when $ \varepsilon
\rightarrow +0; p(\varepsilon) \rightarrow +0 $:
\begin{equation}
\label{5.12} \Lambda(u) \rightarrow + \infty , \hspace{1 cm}
\varepsilon \rightarrow 0; \hspace{1 cm} p(\varepsilon)
\rightarrow 0 .
\end{equation}
Let us investigate the general behaviour of the solution in the
vicinity of a $B$-type singularity, i.e., as $ \Delta(u)
\rightarrow 0 $, imposing a more severe requirement than
(\ref{5.3}) for the equation of state:
\begin{equation}
\label{5.13} p(\varepsilon) \leq \varepsilon/3 .
\end{equation}
Then the following inequalities are valid:
\begin{equation}
\label{5.14} \frac{1}{2 \varepsilon} \leq \drob <
\frac{1}{\varepsilon},
\end{equation}
the equality in the left-hand side (\ref{5.14}) being achieved
only in the case of the ultrarelativistic equation of state ($
\varepsilon = 3p $). Then, restrictions upon $\Je$ follow from
(\ref{5.2}),(\ref{5.14}):
\begin{equation}
\label{5.15} \ln \sqrt{\frac{\varepsilon}{\varepsilon_{0}}} \leq
\Je < \ln \frac{\varepsilon}{\varepsilon_{0}},
\end{equation}
hence by (\ref{5.1}): \footnote {If, instead of (151), we restrict
ourselves to the weaker condition (141), then only a lower bound
upon the energy density will be preserved.}
\begin{equation}
\label{5.16} \frac{1}{\Lambda} < \frac{\varepsilon}
{\varepsilon_{0}} \leq \frac{1}{\Lambda^{2}} .
\end{equation}
But then we obtain directly from  (\ref{4.37}):
\begin{equation}
\label{5.17} \frac{3}{8} L^{2} \Delta^{3} \leq v^{2}_{v} <
\frac{2}{3} \Delta^{2} .
\end{equation}

Using (\ref{5.16}), (\ref{5.17}) and singling out the dominant
parts of the expressions (\ref{4.38}) and (\ref{4.43}) --
(\ref{4.45}) near a B-type singularity, we obtain the asymptotic
estimates $B$:
\begin{equation}
\label{5.18} \Delta \rightarrow 0 \hspace{1 cm}
\begin{array}{ll}
\varepsilon \sim \varepsilon_{0} \Delta^{-\nu};& (\nu \in (1,2])\\
v_{v} \sim \Delta^{\nu}; & \\
H \sim H_{0} \Delta^{-\mu}; & (\mu \in (1,\frac{3}{2}]) \\
v_{u}/v_{v} \sim \Delta^{- \gamma};& (\gamma \in (2,3]) \\
v_{2}/v_{v} \sim \sqrt{2} \cot\Omega \Delta^{-1} &  \\
\end{array}
\end{equation}
The components of the 3-vector of physical velocity$ V^{\alpha} =
dx^{\alpha}/dt$ can be presented in the form:
\begin{displaymath}
V^{1} = \frac{v_{u}/v_{v} - 1}{v_{u}/v_{v} + 1};
\end{displaymath}
\begin{equation}
\label{5.19} V^{2} = \frac{\sqrt{2}e^{-2\beta}} {L^{2} (1+
v_{u}/v_{v})} \frac{v_{2}}{v_{v}} .
\end{equation}
Using the estimates (\ref{5.18}) for the functions near the
singularity, we obtain for the physical velocity vector
components:
\begin{equation}
\label{5.20} \Delta \rightarrow 0: \hspace{1.5 cm}
\begin{array}{ll} V^{1} \rightarrow 1; & \\
V^{2} \sim - \frac{2 e^{-2\beta}}{L^{2}} \cot\Omega \Delta^{\gamma
- 1} & \rightarrow 0 \end{array} .
\end{equation}

Thus a physical singularity occurs on the hypersurface
$u=u_{\ast}$ the energy densities of the plasma and the magnetic
field tend to infinity, the dynamic speed of the plasma as a whole
tends to the speed of light in the PGW propagation direction.
Meanwhile the plasma transverse motion vanishes ($ V^{2}
\rightarrow 0 $), whereas the ratio of the magnetic field energy
density and that of the plasma tend to infinity:
\begin{equation}
\label{5.21} \Delta \rightarrow 0 ; \hspace{1 cm}
\frac{\varepsilon_{H}}{\varepsilon} \sim \alpha^{2} e^{2\beta}
\Delta^{-1} \rightarrow \infty .
\end{equation}

The presence of a $B$-type singularity naturally poses the
question of the applicability of the MHD plasma model in the
vicinity of the hypersurface (\ref{5.8}), i.e., the question of
fulfilling the drift approximation condition (\ref{2.1}).

Note that (\ref{2.1})) is a local condition, i.e., for the MHD
plasma model to be applicable, this condition must be fulfilled in
the comoving FR throughout the whole domain under consideration.
Let us assume that (\ref{2.1}) is fulfilled in the initial state u
$ u \leq 0 $:
\begin{equation}
\label{5.22} \xi_{0} = \frac{\omega}{\stackrel{0}{\omega_{B}}} \ll
1 ,
\end{equation}
where $ \stackrel{0}{\omega_{B}} = \omega_{B|(H =H_{0})}$. The GW
frequency in a frame of reference moving at a speed $v^{i}$, is $
\omega' = k_{i} v^{i} $, where $ {\bf k}$ is the GW wave vector: $
{\bf k} =(-\omega,0,0,\omega)$. Thus $ \omega' = \sqrt{2}
v_{v}\omega $. The scalar $ H $ (\ref{2.16}) is the modulus of the
magnetic field intensity in the comoving FR. So the local value of
the drift parameter $ \xi $, measured in the comoving FR, is
related to the initial value of (\ref{5.22}) by
\begin{equation}
\label{5.23} \xi = \sqrt{2} \frac{v_{v} H_{0}}{H} \xi_{0} .
\end{equation}
Using the asymptotic estimates (\ref{5.18}) and (\ref{5.20}) of
the solution behaviour in the vicinity of a B-type singularity, we
obtain from (\ref{5.23}):
\begin{displaymath}
\Delta \rightarrow 0: \hspace{1 cm} \xi \sim \xi_{0} \Delta^{2\mu}
\rightarrow 0 .
\end{displaymath}

Thus we can conclude: if the drift approximation applicability
condition was initially fulfilled, then in approaching to a
$B$-type singularity the plasma motion is more and more precisely
described by the MHD model. Therefore, within the scope of the
problem in hand, viz., that of motion of initially homogeneous
magnetoactive plasma against the {\em background of the PGW
metric} there are no mechanism available for preventing the
singularity.

We should note an essentially nonlinear nature of the phenomenon
detected: if the initial equations are expanded in a Taylor series
in the small PGW amplitude, $ \beta $, this phenomenon is not
observed, as confirmed, in particular, by an investigation carried
out in Ref. [14]. The reason for this difference of the results
lies in the fact that the governing function of the process under
study is, as can be seen from the solutions to (\ref{4.37}) -
(\ref{4.39}), the function $\Delta^{-1} $, which in the case of a
weak PGW ($ | \beta| \ll 1 $) takes the form:
\begin{equation}
\label{5.24} \Delta^{-1} \approx \frac{1}{ 1 - 2 \alpha^{2}
\beta(u)} .
\end{equation}
The Taylor expansion of this function in powers of $ \beta $
assumes the smallness of the value $ 2 \alpha^{2} \beta $;
however, the parameter $ \alpha $ in a highly magnetized plasma
may prove to be so great that the condition $ 2 \alpha^{2} |
\beta| > 1 $  is fulfilled.

Thus, the discussed phenomenon is a {\em threshold effect} arising
when the PGW amplitude in the compression phase along the axis
$0x^{3}$ reaches the value (\ref{5.11}) and can be interpreted as
{\em a non-linear threshold GW-generation of a
magneto\-hyd\-ro\-dy\-na\-mic shock wave propagating at a
subluminal speed along the PGW propagation direction}. We shall
further call this new class of effects {\em gravimagnetic shock
waves}, or {\em GMSW} for short.

Let us note that, firstly, the presence of a plasma qualitatively
changes the electromagnetic field nature [7],[14]: in the action
of a PGW on a vacuum magnetic field, the solution of the Maxwell
equations is essentially nonstationary (an exact solution is given
in [6]) and, secondly, neither in a fluid, with any equation of
state (exact solutions being given in [15],[16]), nor in a vacuum
magnetic field singularities other than an $ A $-type coordinate
singularity occur. The only and, at the same time, exotic example
is the plasma with a scalar interaction in the case of repulsion
of two identically charged particles (an exact solution is given
in [17]). It should be also noted that in the case of the {\em
ultrastiff equation of state} ($p=\varepsilon$) the solution to
the hydrodynamic equations is nonstationary, i.e., depends on the
variables $u$ and $v$ [16]. The same type of behaviour is detected
for a magnetoactive plasma: as seen from (\ref{5.1}) -
(\ref{5.2}), at $p=\varepsilon$ there is no stationary solution.

The shock wave formation mechanism seems to consist in the
following. A weak GW is known not to interact with a fluid but to
perturb a magnetic field. This in turn causes a drift of the
plasma. Particles with smaller values of the coordinate $x^{1}$
have a greater value of the coordinate $u$, that is why at the
compression stage of the geodesic tube in the direction $0x^{3}: (
\beta >0;\,\, \beta'>0 )$ such particles have a greater speed than
those with larger values of the coordinate $x^{1}$. Therefore, the
backward plasmic layers overtake forward ones and thereby
contribute to the shock wave formation.

Calculating the drift current with the aid of the Maxwell
equations (\ref{3.32}), we find (see also [6],[14]):
\begin{equation}
\label{5.25} J_{3} = \frac{e^{-2\beta}}{4\pi} ( L^{2} F^{3}_{\
v})'= -\frac{H_{0}\sin\Omega} {2\sqrt{2}\pi} \beta' .
\end{equation}
For this reason, in a singular $B$-type state the drift current
density remains finite.

\subsection{Barotropic equation of state}
We shall examine a barotropic equation of state:
\begin{equation}
\label{5.26} p = k \varepsilon , 0 \leq k < 1 .
\end{equation}
Then
\begin{equation}
\label{5.27} \Je = \frac{(1-k)}{(1+k)} \ln
\frac{\varepsilon}{\varepsilon_{0}}
\end{equation}
and thus, according to (\ref{5.2}), (\ref{4.37}) -- (\ref{4.40}),
we obtain an exact solution:
\begin{equation}
\label{5.28} \varepsilon = \varepsilon_{0} \Lambda^{-1+ \kappa} ;
\end{equation}
\begin{equation}
\label{5.29} v_{v} = \frac{1}{\sqrt{2}} L^{\kappa} \Delta^{1 +
\frac{\kappa}{2}} ;
\end{equation}
\begin{equation}
\label{5.30} \frac{v_{u}}{v_{v}} = \Delta^{-2} \left[
\Lambda^{-\kappa} + (\Delta -1)^{2} L^{-2} e^{-2\beta}
\cot^{2}\Omega \right] ;
\end{equation}
\begin{equation}
\label{5.31} H^{2} = \frac{H^{2}_{0}}{\Lambda^{2}} = \left(
\cos^{2}\Omega + L^{2} \Lambda^{-\kappa} e^{2\beta} \sin^{2}\Omega
\right) ,
\end{equation}
where
\begin{equation}
\label{5.32} \kappa = \frac{2k}{1-k} \geq 0.
\end{equation}

In particular, for a nonrelativistic plasma\footnote{The exact
solution for a nonrelativistic plasma was obtained in [18]} ($
p=0, \Rightarrow \kappa=0$) (\ref{5.28})-(\ref{5.31}) imply:
(\ref{5.28})-(\ref{5.31}):
\begin{displaymath}
\varepsilon =\varepsilon_{0} \Lambda^{-1} ; \hspace{1 cm} v_{v} =
\frac{1}{\sqrt{2}} \Delta ;
\end{displaymath}
\begin{displaymath}
\frac{v_{u}}{v_{v}} = \Delta^{-2} \left[ 1+ (\Delta - 1)^{2}
L^{-2}e^{-2\beta}\cot^{2}\Omega \right] ;
\end{displaymath}
\begin{equation}
\label{5.33} H^{2} = \frac{H^{2}_{0}}{\Lambda^{2}} \left(
\cos^{2}\Omega + L^{2} \sin^{2}\Omega e^{2\beta} \right) ;
\end{equation}
and for an ultrarelativistic plasma ($p=\varepsilon/3 \Rightarrow
\kappa =1 $) we obtain from (\ref{5.24}) -- (\ref{5.28}):
\begin{displaymath}
\varepsilon = \varepsilon_{0}\Lambda^{-1} ;\hspace{1.5 cm} v_{v} =
\frac{1}{\sqrt{2}} L \Delta^{3/2} ;
\end{displaymath}
\begin{displaymath}
\frac{v_{u}}{v_{v}} = L^{-2} \Delta^{-2} \left[
\Delta^{-1}+(\Delta -1)^{2} e^{-2\beta} \cot^{2}\Omega \right] ;
\end{displaymath}
\begin{equation}
\label{5.34} H^{2} = \frac{H^{2}_{0}}{\Lambda^{2}} \left(
\cos^{2}\Omega + \Delta^{-1} e^{2\beta}\sin^{2}\Omega \right) .
\end{equation}

The exact solutions (\ref{5.28}) - (\ref{5.31}) confirm the
asymptotic estimates for the general behaviour of plasma near the
singularity (\ref{5.18}).

\subsection{Drift of a nonrelativistic plasma in a weak PGW}
Let us examine the practically important case of a {\em weak PGW}:
\begin{equation}
\label{5.35} |\beta(u)| \ll 1
\end{equation}
and {\em a nonrelativistic plasma} ($k=0$). It is known (see, for
instance, [12]) that $L^{2}\sim O(\beta^{2})$. In view of this
fact, we obtain from (\ref{5.33}) in the first approximation with
respect to $\beta$ (but not with respect to $\alpha^{2} \beta$!):
\begin{displaymath}
\varepsilon=\varepsilon_{0}\Delta^{-1} ;\hspace{1.5 cm}
v_{v}=\frac{1}{\sqrt{2}} \Delta ;
\end{displaymath}
\begin{displaymath}
\frac{v_{u}}{v_{v}} = \Delta^{-2} [1+(2 \alpha^{2} \beta
\cot\Omega)^{2}] ;
\end{displaymath}
\begin{equation}
\label{5.36} H= \frac{H_{0}}{\Delta} ; \hspace{1 cm}
\frac{v_{2}}{v_{v}} = \sqrt{2}(\Delta^{-1} -1)\cot\Omega ,
\end{equation}
where it is necessary to use the expression (\ref{5.24}) for
$\Delta^{-1}$. Then we find from (\ref{5.19}) the nonzero physical
speed components in an explicit form:
\begin{displaymath}
V^{1} = 2\alpha^{2}\beta \frac{1+\alpha_{0}^{2}\beta
\cos2\Omega}{1-2\alpha^{2}\beta(1-\alpha_{0}^{2}\beta)} ;
\end{displaymath}
\begin{equation}
\label{5.37} V^{2} = -\frac{1}{\sqrt{2}}\alpha^{2}_{0}\beta
\sin2\Omega \frac{1-2\alpha^{2}\beta}{1-\alpha^{2}\beta +
2\alpha^{4}_{0}\beta{2}\cos^{2}\Omega} ,
\end{equation}
with $\alpha^{2}_{0} = H^{2}_{0}/4\pi(\varepsilon_{0} + p_{0})$.

It follows from (\ref{5.37}) that in the case of a sufficiently
weak PGW ($2\alpha^{2}|\beta| <1)\, V^{1} >0$ for $\beta >0$ and
$V^{1} <0$ for $\beta <0$, i.e., in the compression phase of of
the geodesic tube in the direction of the $Ox^{3}$ axis, the
plasma drifts as a single whole in the the PGW propagation
direction, whereas in the expansion phase it drifts in the
opposite direction.

Let $\beta(u)$ be a quasiperiodical function with the period $T$ ,
so that
\begin{equation}
\label{5.38} \langle\beta(u)\rangle =0 ,
\end{equation}
where $\langle\cdots\rangle$ denotes averaging over the interval
$\Delta u = u - u_{0}\gg T $ ;
\begin{displaymath}
\vert e^{2\beta(u+T)}-e^{2\beta(u)}\vert \ll\frac{1}{\alpha^{2}} ;
\end{displaymath}
\begin{equation}
\label{5.39} \vert e^{2\beta(u+T/2)}+e^{2\beta(u)}-2\vert \ll
\frac{1}{\alpha^{2}} , --
\end{equation}
i.e., the PGW amplitude is little changed in the course of the
"period" $ T $. Setting in (\ref{5.37}) $\Omega = \pi/2$ and
averaging  $V^{1}$ under the conditions (\ref{5.38})-(\ref{5.39})
over a sufficiently large interval of retarded time, we find:
\begin{equation}
\label{5.40} \langle V^{1}\rangle \sim
 2\alpha^{4}_{0}\langle\beta^{2}(u)\rangle ,-
\end{equation}
thus the average plasma drift proceeds in the PGW propagation
direction and even in a weak GW it is nonzero. At $\Omega = \pi/2$
by (\ref{5.36}) $v_{u}/v_{v} \equiv v^{v}/v^{u} = \Delta^{-2}$,
therefore:
\begin{displaymath}
\frac{dv}{du} = \frac{1}{\Delta^{2}} = \frac{1}{(1 - 2
\alpha_{0}^{2} \beta(u))^{2}}\, ;\Rightarrow
\end{displaymath}
\begin{equation}
\label{5.41} v - v_{0} = \int \limits_{0}^{u} \frac{du'}{(1 - 2
\alpha_{0}^{2} \beta(u'))^{2}} .
\end{equation}
Passing on in (\ref{5.41}) to the $ u \rightarrow u_{\ast} $ under
the condition that the function $ \beta(u)$ be continuous, we get:
\begin{equation}
\label{5.42} u \rightarrow u_{\ast}\, ;\hspace{1 cm} v(u) - v_{0}
\simeq \frac{1}{(2 \alpha^{2} \beta'(u_{\ast}))^{2}}
\frac{1}{(u_{\ast}- u)} \rightarrow \infty
\end{equation}
Eq. (\ref{5.41}) described in an implicit form the trajectory $x
\equiv x^{1} = f(t)$ of an infinitely small volume of plasma. Let
this volume be placed at a point $x_{0}$ prior to PGW arrival,
then $t_{0} = x_{0}$ is the moment of PGW arrival at this point.
The singularity occurs at $u=u_{\ast} =\frac{1}{\sqrt{2}} (t -
x)_{\ast}$. Then it follows from (\ref{5.42}) that in this state
$x_{\ast} \rightarrow +\infty\,; t_{\ast} \rightarrow +\infty $.
Thus, by an external observer's time, a plasmic particle reaches
the singular hypersurface $u=u_{\ast}$ (which the plasma velocity
tends to that of light and the energy density tends to infinity)
for an infinitely long time, and, in doing so, it finds itself at
an infinitely remote point. If the values $|\beta(u)|>
1/2\alpha_{0}^{2}$ are possible in the PGW metric (\ref{4.1}), it
makes no sense to continue the solution of magnetohydrodynamics
beyond the hypersurface $u=u_{\ast}$, since at the shock wave
front all the invariants suffer a secondtype discontinuity.

\section{Source of energy}
Since on a singular hypersurface $u=u_{\ast}$ the energy densities
of the plasma and the magnetic field tend to infinity, while the
velocity of plasma as a whole tends to the speed of light, the
total energy of a MHD shock wave tends to infinity. Thus, as a
result of the drift, the the complete plasma energy grows (indeed,
it grows to infinity), and it is necessary to reveal the nature of
a source of this energy. Note that the MHD equations were solved
against the {\em background of the PGW metric}. In this sense the
PGW is an inexhaustible source of energy for the magnetoactive
plasma; it is precisely for this reason that a singular state is
formed. A singular state originating in the plasma under the
influence of PGW, disturbs the basic assumption (\ref{1}) on the
weakness of the GW-plasma interaction. In a more comprehensive
self-consistent problem with regard to {\em gravitation} the
allowance made for the impact of the shock wave on the PGW should
lead to a PGW amplitude damping to the values
\begin{equation}
\label{5.43} \mbox{max}(|\beta|) < 1/2\alpha^{2}.
\end{equation}
Thus the discovered effect of MHD wave generation by a
gravitational wave (GMSW) can be an effective mechanism of GW
energy absorption. The author is unaware of any other GW energy
absorption mechanism of similar efficiency.

A strict solution to the problem of PGW energy transformation to
the energy of a shock wave is possible only on the basis of a
study of the self-consistent set of the Einstein and MHD
equations. However, a qualitative analysis of the situation may be
performed using a simpler model of the energy balance. Let us
consider the case of strictly transverse PGW propagation
($\Omega=\pi/2$) in a nonrelativisic plasma. Then the energy flux
of magnetoactive plasma should have the direction of PGW
propagation, i.e., along the $0x^{1}$ axis. Let $\beta_{0}(u)$ be
the PGW vacuum amplitude and $\beta(u)$ be its amplitude with
allowance made for absorption in the plasma. From summed energy
flux conservation it follows:
\begin{equation}
\label{5.44} T^{41}(\beta) + t^{41}(\beta) = t^{41}(\beta_{0})\, ,
\end{equation}
where $t^{14}$ is the energy flux of a weak GW in the direction
 $0x^{1}$ (see [11]). Using the solution to the MHD equations for
nonrelativistic plasma (\ref{5.36}) in the case of strictly
transverse PGW propagation, as well as the expressions for the
summed plasma EMT (\ref{2.37}) and weak PGW energy flux [11], we
reduce (\ref{5.44}) to the form (returning to the standard system
of units)
\begin{displaymath}
\frac{\pi G \varepsilon_{0}}{c^{2}}
[\Delta^{-4}(\beta)-1][-\alpha^{2}_{0} e^{2\beta} + \Delta(\beta)]
+ \dot \beta^{2} = \dot \beta_{0}^{2} ,
\end{displaymath}
where $\dot \beta$ means the derivative with respect to the time
$t$. Setting in what follows $ \dot \beta^{2} = \omega^{2}
\beta^{2}\,\\ (\omega$ being the GW frequency), $\alpha^{2}_{0}
\gg 1 $ and $|\beta|\ll 1$ and tending $\Delta$ to zero, we reduce
the latter equation to the form:
\begin{equation}
\label{5.45} \frac{\chi^{2}}{(1-2 \alpha^{2}_{0} \beta)^{4}} +
\beta^{2} = \beta^{2}_{0} \, ,
\end{equation}
where $\chi$ is a dimensionless parameter:
\begin{equation}
\label{5.46} \chi^{2} = \frac{G H^{2}_{0}}{c^{2} \omega^{2}} \sim
\frac{\omega^{2}_{g}}{\omega^{2}}\, ,-
\end{equation}
$\omega^{2}_{g} = 8\pi G {\cal E}_{0}/c^{2}$. The approximation
(\ref{1}) is equivalent to the condition:
\begin{equation}
\label{5.47} \chi^{2} \ll 1.
\end{equation}
In the conditions of strong PGW energy absorption we have
$\beta^{2}\ll \beta^{2}_{0}$, therefore in this case (\ref{5.46})
implies:
\begin{equation}
\label{5.48} 1 - 2\alpha^{2}_{0} \beta \approx
\left(\frac{\chi}{|\beta_{0}|}\right)^{1/2} \Longrightarrow \beta
=  \frac{1}{2\alpha^{2}_{0}} \left[1 -
\left(\frac{\chi}{|\beta_{0}|}\right)^{1/2}\right].
\end{equation}
It follows from (\ref{5.48}) that the conditions for a
sufficiently strong absorption of a weak PGW are:
\begin{equation}
\label{5.49} \alpha^{2} \gg 1 \, ; \hspace{1cm} |\beta_{0}| >
\frac{1}{2\alpha^{2}_{0}} \, \hspace{1cm} |\beta_{0}| > \chi \, .
\end{equation}
The PGW amplitude damping factor $\gamma$ is:
\begin{equation}
\label{5.50} \gamma = \frac{|\beta|}{|\beta_{0}|} =
\frac{1}{2\alpha^{2}_{0}|\beta_{0}|} \left[
1-\left(\frac{\chi}{|\beta_{0}|}\right)^{1/2}\right].
\end{equation}

\section{Astrophysical consequences of GMSW}
There naturally arises a question, whether or not the GMSW exist
in nature, i.e., can the condition for their creation (\ref{5.11})
be realized? In laboratory experiments the above necessary
condition ($2\alpha^{2}|\beta|\geq 1$) is unattainable due to the
smallness of both the gravitational waves in terrestrial
conditions and stable laboratory magnetic fields. For galactic
magnetic fields ($ H \sim 10^{-5}\div 10^{-6}$ Gauss) and
interstellar medium $(\rho \sim 10^{-24}g/cm^{3})\, \to \,
\alpha^{2} \sim 10^{-8}$ ), so GMSW in the interstellar medium do
not arise.

The only possible source of a GMSW can apparently be neutron star
magneto\-sphe\-res on the stage of quad\-ru\-pole neutron
oscillations, as well as on stage of a Supernova. Table 1
represents the results of calculating the parameter
$2\alpha^{2}\langle|\beta|\rangle, \, (\langle|\beta|\rangle$ is
the average amplitude of irradiated GW) for neutron stars
magnetospheres, performing quadrupole oscillations of the
fundamental quad\-ru\-pole mode ($ n=0$). The data placed in the
first four columns are taken (or recalculated) from the book [12].
The average GW amplitude is calculated using standard formulas,
from the average gravitational radiation power, $N$, (see for
instance [11]) under the assumption that the average relative
amplitude of quadrupole oscillations of a neutron star (the
parameter $\langle(\delta R/R)^{2}\rangle^{1/2}$ in the book [12])
is equal to $10^{-6}$. It is also assumed that the neutron star
magnetosphere consists of completely ionized hydrogen. In
addition, the parameter $n_{0}$ (electron number density on the
neutron star surface) was found from the condition $\tau=1$, where
$\tau = \int n\sigma dl$ is the optical thickness of the
magnetosphere. It is assumed also, that the scale of dense
magnetosphere is of the order of the neutron star radius $R$.

Note that the electron concentration values near a neutron star
surface, obtained from the condition that the optical thickness of
the magnetosphere equals unity, is approximately by 4 orders of
magnitude larger than the estimates of $ n_{0} $ from [19],
obtained on the basic of a dimensional analysis of the Maxwell
equations. If we accept Pacini's estimates for $ n_{0} $ [19],
then the GMSW factor will grow by 4 orders as compared with that
of Table 1.
\vskip 12pt
\noindent {\bf Table 1}.\footnote{quantities in parantheses mean
the degrees of 10} {\em The GMSW parameter:
$2\alpha^{2}\langle|\beta|\rangle$ in a neutron star
magnetosphere}
\begin{flushleft}
\begin{displaymath}\hspace{-22pt}
\begin{array}{|l|l|l|l|l|l|l|}
\hline M/M_{\odot} & R & \omega & N/\langle(\delta R/R)^{2}\rangle
& \langle|\beta|\rangle & n_{0} & 2\alpha^{2}\langle|\beta|\rangle
\\ \hline
0,405 & 5,00 & 5249 & 1,2(50) & 2,93(-10) & 5,97(18) & 0,260 \\
\hline
0,682 & 8,42 & 2,02(4) & 2,9(53) & 4,20(-9) & 3,54(18) & 3,32 \\
\hline 0,677 & 12,60 & 8987 & 7,0(52) & 1,64(-9) & 2,36(18)
& 3,670 \\
\hline 1,954 & 9,99 & 1,66(4) & 1,6(55) & 1,41(-8) & 2,99(18)
& 30,0 \\
\hline
\end{array}
\end{displaymath}
\end{flushleft}
\noindent {\small $M/M_{\odot}$ is the neutron star mass related
to the Solar mass; $R$ is the star radius in km.; $\omega$ is the
gravitational radiation frequency in sec$^{-1}$; $N/\langle(\delta
R/R)^{2}\rangle$ is the star's gravitational radiation power in
erg/sec; $\langle|\beta|\rangle$ is the GW average amplitude;
$n_{0}$ is the concentation of electrons on the stellar surface in
$cm^{-3}$.}

\vskip 0.5cm

For a star with a mass of 0,682 $M_{\odot}$ we get from Table 1:
$\chi = 1,33\cdot 10^{-11}; \, \chi/\langle |\beta_{0}|\rangle
=0,317$. According to (\ref{5.50})) we find the GW damping factor:
$\gamma=0,206$. Thus, the GW amplitude in this case is damped by a
factor of 5 and the gravitational radiation power by a factor of
25. For a star with a mass of $1,954 M_{\odot}$ \, $\chi =
1,42\cdot 10^{-11}; \, \chi/\langle|\beta_{0}|\rangle = 1,01\cdot
10^{-3} \rightarrow \gamma = 0,032$, i.e., in this case the GW
amplitude can be diminished by a factor of 30 times and the GW
power by a factor of 900! So in neutron star magnetospheres there
exist {\em necessary} conditions for GMSW excitation.

If the magnetic field of a neutron star is described as that of a
dipole, then the geographic angle $\Theta$ (counted relative to
the magnetic equator) will be connected with the above angle
$\Omega$ by the relation $\Omega = \pi/2 - \Theta $. Therefore the
GMSW excitation condition depends on the angle $\Theta$:
\begin{displaymath}
\label{7.1} \sin^{2}\Theta < 1 - \frac{1}{2\alpha^{2}_{0}|\beta|}.
\end{displaymath}

Thus, in the magnetosphere of a neutron star (or a Supernova) a
GMSW can be excited in the region of the magnetic equator,
similarly to pulsars with a knife-like radiation pattern. In this
region, as was demonstrated by the above examples, the
gravitational radiation can be absorbed practically completely by
the excitation of shock waves. A neutron star of such type should
radiate gravitational waves only from its magnetic poles,
similarly to pulsars with a pencil-like radiation pattern. In this
case the probability of observing a GW sources can be sharply
dropped. However, suddenly the GMSW open another way of observing
gravitational waves. If such a shock wave is formed and takes off
the magnetosphere, it will carry with itself (at a subluminal
speed) superstrong magnetic fields into the interstellar space.
The interaction of cosmic plasma with such magnetic fields can
lead to the anomalous electromagnetic phenomena in the radio and
optical spectral ranges. It should be stressed that there is no
other mechanism able to accelerate a shock wave to subluminal
velocities.

Thus, the GMSW phenomenon may displace the "centroid" of GW
ex\-pe\-ri\-ments from direct GW detection to optical observations
of the Supernovae and their remainders. However, to be quite
certain that the GMSW do exist in neutron star magnetospheres and
to know the observational manifestations of this phe\-no\-me\-non,
it is necessary to solve a number of problems:

\begin{enumerate}
\item  To solve the self-consistent problem of GW propagation in a
homogeneous plasma with an embedded magnetic field and calculate
the damping dec\-re\-ment;
\item  To clarify the effects on the this phenomenon from plasma
inhomogeneity and external gravitational fields;
\item To clarify the effect on this phenomenon from the curvature of magnetic lines
of force;
\item To investigate optical manifestations of GMSW.
\end{enumerate}

We intend to study these problems in future papers.

\end{document}